\documentclass[prl,aps,onecolumn,superscriptaddress,showpacs,floatfix]{revtex4}
\usepackage{epsfig,graphicx,latexsym}
\usepackage{amssymb,amsmath}
\newcommand{\eeta}{{\boldsymbol{\eta}}}
\newcommand{\xxi}{{\boldsymbol{\xi}}}
\begin{document}

\title{\bf\noindent Relaxation  of the thermal Casimir force between net neutral plates containing Brownian charges}

\author{David S. Dean}
\affiliation{Universit\'e de  Bordeaux and CNRS, Laboratoire Ondes et
Mati\`ere d'Aquitaine (LOMA), UMR 5798, F-33400 Talence, France}
\author{Rudolf Podgornik}
\affiliation{Department of Physics, University of Massachusetts, Amherst, MA, USA}\affiliation{Department of Theoretical Physics, J. Stefan
Institute, SI-1000 Ljubljana, Slovenia} 
\affiliation{Department of Physics, Faculty
of Mathematics and Physics, University of Ljubljana, SI-1000
Ljubljana, Slovenia}
\begin{abstract}
We investigate the dynamics of thermal Casimir interactions between plates described within a {\sl living conductor } model, with embedded mobile anions and cations, whose density field obeys a stochastic partial differential equation which can be derived starting from the Langevin equations of the individual particles. This model describes the thermal Casimir interaction in the same way that the fluctuating dipole model describes van der Waals interactions.  The model is analytically solved in a
 Debye-H\"uckel-like  approximation. We identify several limiting dynamical regimes where the time dependence of the thermal Casimir interactions can be obtained explicitly. Most notably we find a regime with diffusive scaling, even though the charges are confined to the plates and do not diffuse into the intervening space, which makes the diffusive scaling difficult to anticipate and quite unexpected on physical grounds.
\end{abstract}
\maketitle

\section{Introduction}
The quantum Casimir effect is the first and most famous example of what are commonly referred to as the {\sl fluctuation induced interactions} \cite{Dalvit}. The idea behind the original calculation by Casimir is quite straightforward to describe in principle but is at the same time mathematically quite technical and, in addition, rather subtle to interpret physically \cite{Casimir}. The basic point  is nevertheless clear: the presence of two conducting plates modifies the ground state energy of the electromagnetic field \cite{Mostepanenko}. The ground state, or zero point energy, then has a finite component which depends on the plate separation, but also a divergent contribution which is fortunately independent of the plate separation, thus leading to a finite force. A physically measurable force is therefore derivable from ${1\over 2}\hbar \omega$, the quantum ground state energy. This is somewhat embarrassing as physics students are often  confidently told that the ground state energy can be thrown away in condensed matter and quantum field calculations, since only differences in energy should count. The question as to whether zero point energy has physically measurable consequences is an important one in cosmology as it is a possible candidate for dark energy \cite{Elizade}.  

Another query arising from Casimir's computation, based on zero point energy, is why is the effect then usually referred to as a fluctuation induced interaction? One computes the ground state energy $E_0$ of a system of harmonic oscillators which, up to an infinite but constant term, is a function of the plate separation $L$ and the ensuing force is simply $f=-{\partial E_0\over \partial L}$. What is fluctuating here? The uncertainty principle tells us the local electromagnetic field fluctuates but it is not clear physically why such fluctuations should give rise to a force. The fluctuation aspect of Casimir interactions and their generalization, the van der Waals forces, has been worked out particularly in detail by Lifshitz. In his landmark  paper, Lifshitz used Rytov's fluctuating electrodynamics to derive the average stress tensor in a planar geometry bounded by two dielectric and/or conducting bodies \cite{Lifshitz1}. This represents the {\em field view} of fluctuation induced electromagnetic forces, in distinction to the {\em matter view} where these interactions are caused by correlated dipolar fluctuations, yielding the long range van der Waals interactions. It is thus natural and unsurprising to call van der Waals interactions fluctuation induced interactions \cite{par2006}. Indeed, Schwinger was unhappy about the  zero-point energy interpretation of the Casimir interaction to such an extent that this led him to formulate the effect in terms of the currents and sources alone \cite{Milton}. Physically, and in analogy with the van der Waals dipolar fluctuation case, this leads to the realization that a conductor has free charges which move in such a way to effectively impose conducting boundary conditions in Maxwell's equations and the Casimir effect is thus due to the correlation of the charge fluctuations between the two plates in just the same way as van der Waals forces are due to correlations of the dipole fluctuations.  Even with these deep insights into the nature of the Casimir effect, and van der Waals and fluctuation induced interaction in general, the discussion about zero point energy versus charge and current fluctuations continues unabated to this day \cite{jaf2005,cug2012}. 

Recently the representation of dielectric bodies in terms of thermalized dipoles has been exploited to study the classical high temperature behavior of  van der Waals  forces \cite{dea2012}. The dipole field representation is also useful because it allows one to study the dynamics of how the van der Waals interactions evolve in time, as correlations between two initially uncorrelated dielectric slabs set in (the fluctuations of the force in equilibrium may also be studied \cite{dea2013}). The input into such a theory is a local Langevin dynamics for the dipole field. The classical equilibrium thermal interaction between two semi-infinite planar slabs of area $A$ separated by a distance $L$ is given by \cite{par2006}
\begin{equation}
f= -{k_BT A H_{eq}\over L^3},
\end{equation}
where $H_{eq}$ is the Hamaker coefficient measured in units of $k_BT$. Two initially uncorrelated slabs 
are then found to have a time dependent interaction of the form \cite{dea2012}
\begin{equation}
f= -{k_BT A H(t)\over L^3}, \label{dipd}
\end{equation}
where $H(t)$ is a time dependent Hamaker coefficient with $H(0)=0$, corresponding to zero force for initially uncorrelated slabs, and $\lim_{t\to\infty} H(t) = H_{eq}$.  One should note however that the factorization of the temporal and spatial dependence in Eq. (\ref{dipd}) is particular to planar geometries and does not have straightforward analogies in other geometries.

Few other results are known about the temporal evolution of fluctuation induced forces. Most results on the electromagnetic Casimir effect concern steady state out of equilibrium situations where the temperatures of the bodies in interaction are taken to be different \cite{dor1998,pit2005,pit2008,bim2009,Emig1,Emig2,mes11,rod11}, these systems are thus in non-equilibrium but steady states. In addition, the quantum Casimir-Polder interaction between atoms and surfaces has been studied in a number of out of equilibrium  contexts \cite{shr03,buh08,mes10,beh11}. Here one can also study the evolution of a quantum state of the system which is not a stationary state, as well as the Casimir friction effect due to the motion of the atom \cite{Mkrtch,Kardar1}. These studies  are more closely related to the one that will be presented here in that we study the evolution of the Casimir force in an initially out off equilibrium state to its equilibrium value.

In soft matter type systems, where a multitude of fluctuation induced interactions exist,  there are only analytical results for simple Gaussian models of binary mixtures with stochastic non-conserved order parameter (model A) dynamics. Here the temporal and $L$ dependence of the force are mixed via a diffusive scaling dependence $L^2/t$ \cite{gamcout,deacout}. In principle, near critical binary liquid mixtures are ideal systems to observe out of equilibrium Casimir interactions, due to the critical slowing down of dynamics as a critical point is approached. However, in such systems, hydrodynamics needs to be taken into account and the problem becomes very difficult, even in the regime of low Reynolds number where flows can be treated as Stokes flows. The first results for such systems have been obtained only very recently and require quite sophisticated numerical techniques \cite{fur2013}. 
As is the case for the electromagnetic Casimir effect, the critical Casimir effect out of equilibrium
can also be studied via the induced drag and diffusion on insertions which interact with the fluctuating field \cite{dem2011,fur2013}

In this paper we will examine the interaction between two {\em model conducting surfaces} that contain mobile cations and anions interacting via the three dimensional Coulomb interaction and whose dynamics is governed by a Langevin equation. This model describes 2D electrolyte layers or thin colloidal layers when hydrodynamic effects can be neglected, but can be also viewed as a generic classical model of 2D conductors. The dynamics of the system is treated analogously to the Debye-H\"uckel (DH) approximation, where fluctuations of the local density of anions and cations about the average density is taken to be small. The statics of this model in the DH approximation has been studied in Refs. \cite{bue2005,jan2005} and the model is called the {\sl living conductor model} by the authors of \cite{jan2005}.  The motivation for these studies was in fact how to better understand the role of charge fluctuations in conductors and their effect on the Casimir interaction in order to understand the extrapolation of the quantum Casimir effect, based on perfectly conducting boundary conditions, to the high temperature limit. The choice of terminology living conductor was chosen by the authors of \cite{jan2005} to emphasize the
presence of real charge distributions in the plates as opposed to conducting boundary conditions imposed on the electromagnetic field. 

In this model, and under the DH approximation, we will study how the Casimir force between two initially uncorrelated plates evolves in time. The basic model consists of diffusing ions confined to both plates, each plate representing a confined 2D electrolyte. The electrolytes have a screening length beyond which electrostatic interactions are weak. For plate separations much larger than this screening  length the equilibrium  interaction has a universal, perfect conductor, limit. For separations smaller than the screening length, the interaction is non-universal and depends explicitly on the screening length. We will present analytical results for the evolution of the Casimir interaction in both the long range and short range regimes. The temporal evolution for the force is surprisingly rich and is quite different in the short and long distance regimes. In the long distance universal regime  we shall also see that the temporal evolution of the force is quite different to that of the dipole model studied before \cite{dea2012}, despite the fact that the same universal static limit can be realized in the two models. 

The paper is organized as follows.  We first describe the basic model in terms of its statics and dynamics and also give a simple and  explicit expression for the force between the two plates in terms of the charge densities of the system. The dynamics of the system is then solved by expanding the density fields  about their mean values to first order, in terms of statics this approximation corresponds to the DH approximation.  Expressions are then given for the Laplace transform of the time dependent force between two initially nonconducting plates which pass through an insulating to conducting transition. From this expression we can easily extract the equilibrium  value of the force. We identify two static limits, a long distance universal limit, where the thermal Casimir force is independent of the microscopic model for the charges in the plates, and a short distance non-universal regime where the scaling of the force with separation changes and a dependence on the microscopic parameters of the model appears. There then follows a rather technical 
passage where we carry out an asymptotic analysis of the Laplace transform of the force to extract its short and late time behaviors in both the universal and non-universal regimes.  We conclude by discussing the difference between the  results found here and those in other studies of the relaxation of  thermal Casimir forces and further perspectives in this line of study.

\section{The living conductor model and its dynamics}

\begin{figure}[t!]\begin{center}
	\begin{minipage}[b]{0.412\textwidth}\begin{center}
		\includegraphics[width=\textwidth]{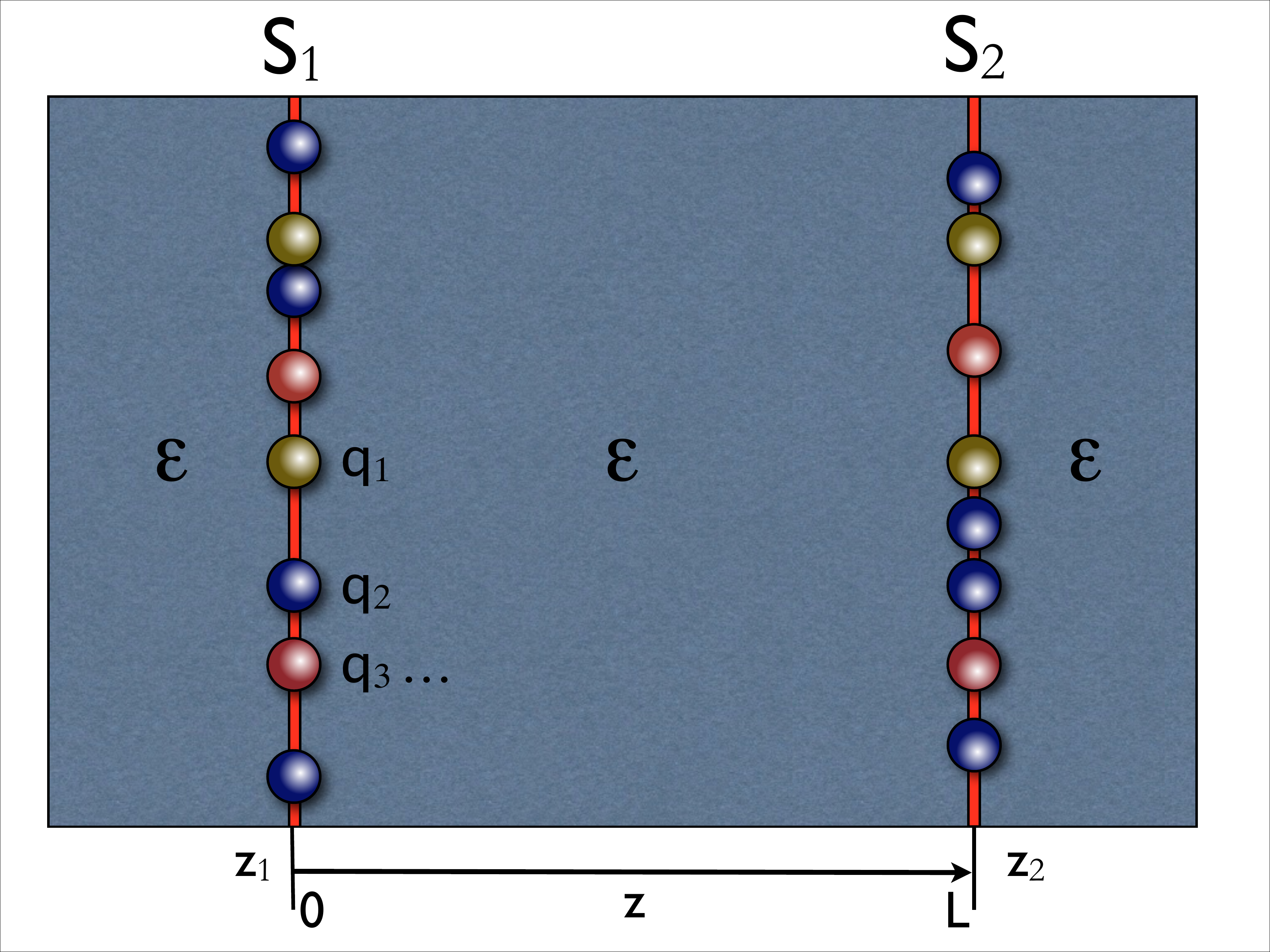}
	\end{center}\end{minipage} 
\caption{(Color online) Schematic of system studied - Brownian anions and cations are contained in two parallel plates separated by distance $L$ in a medium of dielectric constant $\epsilon$}
\label{schema}
\end{center}\end{figure}
We consider two parallel plates $S_1$ and $S_2$ of area $A$ separated by distance $L$ in the direction we denote by $z$. A schematic of the system is shown in Fig. (\ref{schema}).
We set $S_1$ at $z_1=0$ and $S_2$ at $z_2=L$. In the problem we consider cations/anions of several types denoted by $\alpha$. The ions of type $\alpha$ have a charge $q_\alpha$ and are in the plate $S_\alpha$ ($S_\alpha$ = $S_1$ if the ions of type
$\alpha$ are in the plate $1$ etc.). Furthermore the ions are Langevin particles with diffusion 
constant $D_\alpha$. Each particle is thus subjected to a thermal white noise plus the electric field
generated by the other particles (including those not in the same plate). The density field for the particles obeys a stochastic partial differential equation which can be derived starting from the Langevin equations of the individual particles \cite{dea1996} or, as was originally done, from phenomenological arguments \cite{kaw1994}. For a system of different particle types the derivation of \cite{dea1996} can be easily extended to give
\begin{equation}
{\partial\rho_\alpha({\bf x})\over \partial t} = {D_\alpha\over T}\nabla_{||}\cdot  \rho_{\alpha}\nabla_{||}{\delta F\over \delta \rho_{\alpha}} + \nabla_{||}\cdot \sqrt{2 D_\alpha\rho_\alpha} \eeta_\alpha({\bf x})
\end{equation}
where ${\bf x}$ is the two dimensional coordinate in the plane of the plates and $\nabla_{||}$ is the corresponding 2D gradient operator. The functional $F$ is an effective free energy functional for the density fields  given by
\begin{equation}
F = {1\over 2}\sum_{\alpha\beta} \int_{S_\alpha\times S_\beta} d{\bf x} d{\bf x}' 
q_\alpha q_\beta \rho_{\alpha}({\bf x})\rho_{\beta}({\bf x'})G({\bf x}-{\bf x}',z_\alpha-z_\beta) + T\sum_\alpha  \int_{S_\alpha}  d{\bf x}
\rho_\alpha({\bf x})\ln\left(\rho_\alpha({\bf x})\right).\label{fe}
\end{equation}
The first term above is the electrostatic energy of the system and $G$ is the Green's function obeying 
\begin{equation}
\epsilon\nabla^2 G({\bf x}-{\bf x}',z-z') = -\delta({\bf x}-{\bf x}')\delta(z-z')
\end{equation}
with $\epsilon$ the background dielectric constant of the system.  The second term in the functional $F$ corresponds to the entropic contribution $-TS$ due to the ionic distributions. 

The term $\eeta_\alpha$ is spatio-temporal Gaussian white noise with correlation function
\begin{equation}
\langle \eta_{\alpha i}({\bf x},t)\eta_{\beta j}({\bf x}',t')\rangle = \delta_{\alpha\beta}\delta_{ij}\delta({\bf x}-{\bf x}')\delta(t-t').
\end{equation}
Note that the diffusive dynamics above  conserves the total particle number of each species, in what follows we assume that each plate is electro-neutral.
The force between the two plates as
a functional of the densities $\rho_\alpha$ is then given by
\begin{equation}
f= -{1\over 2}\sum_{\alpha\beta: S_\alpha = S_1,\ S_\beta=S_2} \int_{S_\alpha\times S_\beta} d{\bf x} d{\bf x}' 
q_\alpha q_\beta \rho_{\alpha}({\bf x})\rho_{\beta}({\bf x'}){\partial \over\partial L}G({\bf x}-{\bf x}',L),
\end{equation}
where only interactions between charges on different plates contribute to the normal force between the plates. 

Finally we need to specify the initial conditions from which the system evolves. Quickly moving the plates in close proximity and then observing the evolution of the force seems difficult to achieve. A different initial condition would be to start with plates which initially have no free charges and then pass through an insulator/conductor-like transition by tuning an external parameter such as the temperature. The initial condition we shall consider is therefore one where the local charge density is initially zero in both plates. More specifically, if the system is initially made up of bound ion cation pairs, we assume that the initial density of pairs is uniform and the corresponding ionic distributions upon unbinding are thus initially uniform.

\section{Resolution of dynamics in the Debye-H\"uckel approximation}
In order to proceed we expand the density field for each species about its mean value, writing
\begin{equation}
\rho_{\alpha}({\bf x}) = {\overline \rho}_{\alpha} + n_{\alpha}({\bf x}),
\end{equation}
the equation of motion for the density fluctuations $n_\alpha$ is given by
\begin{equation}
{\partial n_\alpha({\bf x})\over \partial t} = {D_\alpha}\nabla^2_{||}n_{\alpha}({\bf x})
+{D_{\alpha}\over T}   \nabla_{||}\left(q_\alpha(\overline\rho_\alpha+n_\alpha({\bf x}))\sum_\beta\int_{S_\beta} d{\bf x}'\nabla_{||}G({\bf x}-{\bf x}', z_\alpha-z_\beta) q_\beta n_\beta({\bf x}')\right)
  + \nabla_{||}\cdot \sqrt{2 D_\alpha(\overline \rho_\alpha+ n_\alpha)} \eeta_\alpha({\bf x},t),
\end{equation}
where we have assumed that each plate is electroneutral, i.e. $\sum_{\alpha: S_\alpha=S_{1/2}} q_\alpha\overline \rho_\alpha=0$. The dynamical equivalent of the DH approximation 
amounts to keeping only terms which are first order in $n_\alpha/\overline{\rho}_\alpha$. In the second noise term this amounts to keeping only the term $\overline\rho_\alpha$, as the first linear term has average $0$ and will only enter corrections quadratically. The dynamics in this approximation becomes equivalent to model B conserved dynamics employed in the study of phase ordering kinetics
\cite{bray}. 

The effective equation thus becomes linear and can be written as
\begin{equation}
{\partial {\bf n}\over \partial t} = -{1\over T}RA{\bf n} + \xxi,
\end{equation}
where ${\bf n}$ is the vector with components $n_\alpha$. By $RA$ we denote the composition of operators. The noise $\xxi$ has correlation function
\begin{equation}
\langle\xi_{\alpha}({\bf x},t) \xi_{\beta}({\bf x}',t)\rangle = -2\delta(t-t')\delta_{\alpha\beta} D_\alpha\overline{\rho}_\alpha\nabla_{||}^2\delta({\bf x}-{\bf x}')
=2 \delta(t-t')R_{\alpha\beta}({\bf x}-{\bf x}').
\end{equation}
From this the dynamical operator $R$  can be read off as
\begin{equation}
R_{\alpha\beta} = -\delta_{\alpha\beta}D_\alpha\overline{\rho}_\alpha\nabla_{||}^2
\end{equation}
and we also find that
\begin{equation}
A_{\alpha\beta}= {T\delta_{\alpha\beta}\delta({\bf x}-{\bf x}')\over \overline{\rho}_{\alpha}} + q_\alpha q_\beta 
G({\bf x}-{\bf x}',z_\alpha-z_\beta).\label{aeq}
\end{equation}
It is easy to see that one has 
\begin{equation}
A_{\alpha\beta}n_\beta = {\delta F_{DH}\over \delta n_{\alpha}}
\end{equation}
where $F_{DH}$ is the DH approximation to the free energy functional in Eq. (\ref{fe}) where it is expanded to second order in ${n_\alpha/\overline{\rho}_\alpha}$. The equal time correlation function $C_{\alpha\beta}({\bf x}-{\bf x}',t) = \langle n_\alpha({\bf x},t)n_\beta({\bf x}',t)\rangle$ is straightforward to compute and we find (in operator notation where we suppress the spatial dependence of the operators for notational succinctness)
\begin{equation}
C(t) = \exp(-{t\over T}RA)C(0)\exp(-{t\over T}AR) + TA^{-1}\left(1-\exp(-2{t\over T}AR)\right),
\end{equation}
where $C(0)$ is the initial correlation function. In the case considered here we start with an
initial insulator condition with $n_\alpha=0$ in both plates, this means that the initial distribution of each charge type is uniform at the moment where the dynamics begins. We therefore set $C(0)=0$.
The average force as a function of time is given by
\begin{equation}
\langle f(t) \rangle= -{1\over 2}\sum_{\alpha\beta: S_\alpha = S_1,\ S_\beta=S_2} \int_{S_\alpha\times S_\beta} d{\bf x} d{\bf x}'  q_\alpha q_\beta C_{\alpha\beta}({\bf x}-{\bf x}',t){\partial \over\partial L}G({\bf x}-{\bf x}',L).
\end{equation}
Here the only terms present are from charge distributions in different plates so we have
\begin{equation}
\langle f(t) \rangle= -{1\over 2}\sum_{\alpha\beta: S_\alpha = S_1,\ S_\beta=S_2} \int_{S_\alpha\times S_\beta} d{\bf x} d{\bf x}'  q_\alpha q_\beta  \left[TA^{-1}\left(1-\exp(-2{t\over T}AR)\right)\right]_{\alpha\beta}({\bf x}-{\bf x}'){\partial \over\partial L}G({\bf x}-{\bf x}',L).
\end{equation}
As seems to be generic with the dynamics of Gaussian fluctuation induced forces \cite{deacout,dea2012}, the analysis is greatly simplified by working with the Laplace transform of the average time dependent force
\begin{equation}
\tilde f_a(s)=\int_0^\infty \exp(-st) \langle f(t) \rangle.
\end{equation}
This gives in operator notation
\begin{equation}
\tilde f_a(s)= -{T\over 2s}\ {\rm Tr} {\partial A\over \partial L} (A + {1\over 2}s TR^{-1})^{-1},
\end{equation}  
where we have used Eq. (\ref{aeq}) in which only the second term on the right hand side depends on $L$. We can now write, as in previous studies \cite{deacout},
\begin{equation}
\tilde f_a(s)= -{T\over 2s}\ {\rm Tr} {\partial A_s\over \partial L} A_s^{-1},
\end{equation} 
where $A_s = A + {s\over 2}TR^{-1}$ as the second term dependent on $s$ is independent of $L$. Note that this 
result is formally similar to that found in \cite{deacout} for general Gaussian fluctuating  fields confined between (and not in) the surfaces, however the physical models and assumptions made are somewhat different. 

Finally  we can write
\begin{equation}
 \tilde f_a(s)= -{T\over 2s}{\partial\over \partial L} {\rm Tr}\ln(A_s) = {T\over s}{\partial\over \partial L} \ln(Z_s),
\end{equation}
 where we have introduced the functional integral
 \begin{equation}
Z_s = \int d[{\bf n}] \exp\left(-{\beta\over 2} \sum_{\alpha\beta}\int_{S_\alpha\times S_\beta}   d{\bf x} d{\bf x}' n_\alpha({\bf x}) A_{s\alpha\beta}({\bf x},{\bf x}')  {n}_\beta({\bf x}')\right)
\end{equation}
The functional integral is easily evaluated by introducing a scalar field $\phi$ which decouples the non-local Coulomb interaction and, up to a factor of $i$, corresponds to the electrostatic field, whereby up to an overall constant
\begin{eqnarray}
&& Z_s =\nonumber \\
&& \int d[{\bf n}]d[\phi]\ \exp\left(-{\beta\over 2}\sum_\alpha \int_{V_\alpha} d{\bf x}\left[ {T\over{\overline \rho}_\alpha} n_\alpha({\bf x})^2 -{sT\over 2 D_\alpha{\overline\rho}_\alpha } n_{\alpha}({\bf x}) \nabla_{||}^{-2} n_{\alpha}({\bf x}) \right] + i\beta \sum_\alpha \int_{S_\alpha} d{\bf x} \phi({\bf x}) q_\alpha n_{\alpha}({\bf x}) -{\beta \over 2} \int d{\bf r}\epsilon  [\nabla \phi({\bf r})]^2\right).\nonumber \\
\end{eqnarray}
Note that in the above functional integral the last term in the exponential is a volume integral over all space, where as the other integrals are surface integrals over the planes containing the charge distributions. The integral over the fields $n_\alpha$ can be performed in terms of their Fourier transforms  with respect to the in plane coordinates ${\bf x}$. This yields
\begin{eqnarray}
Z_s &=& \prod_{{\bf k}} \int d[\phi({\bf k},z)] \exp\left(-{\beta\epsilon  \over 2} \int dz {d\phi(-{\bf k},z)\over dz} {d\phi({\bf k},z)\over dz} + k^2 \phi(-{\bf k},z)\phi({\bf k},z)\right. \nonumber \\ 
&-&{\beta \epsilon\over 2} \left.
\phi(-{\bf k},0)\phi({\bf k},0) m_1(k,s) -{\beta \epsilon\over 2} 
\phi(-{\bf k},L)\phi({\bf k},L) m_2(k,s) \right)\label{fis}
\end{eqnarray}
and where the $m_i(k,s)$ are $k$ and $s$ dependent inverse effective Debye lengths for the ions in each plate given by 
\begin{equation}
m_i(k,s) = \sum_{\alpha:S_\alpha = S_i}  {{\overline\rho}_\alpha q_\alpha^2\over\epsilon  T}  {1\over 1 + {s\over 2D_\alpha k^2}},
\end{equation}
and in the static limit, where $s=0$, they are the usual static inverse Debye lengths for systems restricted to two dimensions. 

The functional integrals in Eq. (\ref{fis}) having surface terms  can be evaluated using a variety of path integral or functional integral techniques \cite{casoft}. This gives an effective free energy $F_s = -\ln(Z_s)/T$ which, up to bulk and surface terms independent of the separation of the plates, is given by
\begin{equation}
F_s = {TA\over 4\pi}  \int kdk \ln\left(1 - {m_1(k,s) m_2(k,s) \exp(-2kL)\over (2k+m_1(k,s))(2k+m_2(k,s))}\right),
\end{equation}
which gives the Laplace transform of the time dependent force to be 
\begin{equation}
\tilde f_a = -{TA\over 2\pi s}  \int k^2dk {m_1(k,s) m_2(k,s) \exp(-2kL)\over (2k+m_1(k,s))(2k+m_2(k,s))- m_1(k,s) m_2(k,s) \exp(-2kL)}.\label{flt}
\end{equation}
This is the main result of this paper. In what follows we analyze this result, and notably carry out the inversion of the Laplace transform to deduce the temporal evolution of the force.  
\section{Static limit}
In the static limit $s\to 0$ the equilibrium force is given via the pole at $s=0$, we thus find
\begin{equation}
\langle f\rangle_{eq}=-{TA\over 2\pi }  \int k^2dk {m_1 m_2 \exp(-2kL)\over (2k+m_1)(2k+m_2)- m_1 m_2 \exp(-2kL)}.,
\end{equation}
where 
\begin{equation}
m_i = \sum_{\alpha:S_\alpha = S_i}  {\overline{\rho}_\alpha q_\alpha^2\over\epsilon  T}.  
\end{equation}
The $m_i$ represent the inverse of  length scales $l_i$ beyond which  the interaction between
ions in the same plates are screened. These fix length scales which determine the Casimir
interaction between the plates, the large distance limit where $L\gg l_i$ and  the short distance 
limit where $L \ll l_i$.

\subsection{Long distance universal limit}

In the limit where $Lm_i\gg 1$ we find that
\begin{eqnarray}
\langle f\rangle_{eq}&=&-{TA\over 16\pi L }  \int u^2du{ m_1 m_2 \exp(-u)\over (u+m_1L)(u+m_2L)- m_1 m_2 L^2\exp(-u)}\nonumber \\
&\approx& f_{uni}= -{TA\over 8\pi L^3}\zeta(3).
\end{eqnarray}
This  is the universal limit for conductors in the classical limit. Its value is half that obtained by taking the 
high temperature limit of the ideal conductor Casimir limit, however it is the same result as that given 
by Lifshitz theory. This subtlety of the high temperature limit of the Casimir effect was in fact what 
inspired the authors of refs \cite{bue2005} and \cite{jan2005} to study this {\sl living conductor }model. 
The finite distance corrections are easily computed to first order giving
\begin{equation}
\langle f\rangle_{eq} \approx f_{uni}\left[ 1- {3\over L}({1\over m_1}+ {1\over m_2})\right],\label{eqpc}
\end{equation}
which agrees with the correction given in \cite{jan2005} where the case $m_1=m_2$ is considered. 
\subsection{Short  distance non-universal limit}
Now in the opposite limit where the plate separation  is smaller than both  their individual screening lengths, $Lm_i \ll1$, we find
\begin{equation}
\langle f\rangle_{eq}= -{TAm_1m_2\over 16\pi L},\label{mpstat}
\end{equation}
giving a completely different scaling form when compared with the long distance limit. We first note that the interaction force in this limit is not universal, but depends on $m_1, m_2$. The inter-surface distance scaling can be understood as follows: monopolar charge fluctuations between two point-like charge distributions give a free energy that goes as the inverse first power of the separation squared, just as the dipolar fluctuations scale as the inverse third power of the separation squared; a Hamaker-like summation of the inverse second power forces, distributed uniformly over two apposed planar surfaces, then yields a net force that scales as the inverse first power of the separation between the planar layers. 
A crucial point leading to Eq. (\ref{mpstat}) is however that the monopolar charge fluctuations are correlated leading to an attractive force.
\section{Early time relaxation of the force}
The very short time behavior of the force is easy to extract. In this limit we assume that for all $\alpha$ that $L^2/D_\alpha t\ \sim  sL^2/D_\alpha \gg1$. This means that we can take $m_i(k,s)\ll 1$ in Eq. (\ref{flt}) and we find
\begin{equation}
\tilde f_a \approx -{TA\mu_1\mu_2\over 8\pi s^3} \int  k^4dk  \exp(-2kL),
\end{equation}
where $\mu_i = \sum_{\alpha:S_\alpha = S_i}  {2D_\alpha{\overline \rho}_\alpha q_\alpha^2\over\epsilon  T}$
and thus
\begin{equation}
\tilde f_a \approx -{3TA\mu_1\mu_2\over 32 \pi s^3 L^5}. 
\end{equation} 
The inversion of the Laplace transform in this large $s$ limit corresponds to short times and we 
find
\begin{equation}
\langle f(t) \rangle  \approx-{3TA\mu_1\mu_2 t^2\over 64 \pi  L^5}. \label{st}
\end{equation} 
As is seen for systems of interacting dipoles, the growth of the force as $t^2$ is a signal of the fact that
it is correlation induced and thus in a sense is second order in time. However, the short time scaling with $L$ is quite different to the dipole case as is seen in Eq (\ref{dipd}). We also see straight away that, in contrast 
with the dipole case, the time scales for the evolution of the force are dependent on $L$. This is somewhat surprising as, in common with the dipole case, the effective interaction is again generated 
by an instantaneous Coulomb interaction (in the dipole case the interaction between the partial charges
on the dipoles is of course Coulombic). 

\section{Late time relaxation of the force}
Here we consider how the force evolves at late times to its static value. As there are two distinct regimes
for the statics determined by the relative values of $L$ and $l_i$ we need to consider them separately.

\subsection{Late time evolution in the short distance regime}
As in the static case we consider the change of variables $k=u/L$ and we recall that $s\sim 1/t$ so that 
late time, or large $t$, dynamics corresponds to small $s$. The dynamical inverse screening lengths 
behave as
\begin{equation}
m_i(k,s) = \sum_{\alpha:S_\alpha = S_i}  {{\overline\rho}_\alpha q_\alpha^2\over\epsilon  T}  {1\over 1 + {sL^2\over 2D_\alpha u^2}} \sim  \sum_{\alpha:S_\alpha = S_i}  {\overline{\rho}_\alpha q_\alpha^2\over\epsilon  T}  {1\over 1 + {L^2\over 2D_\alpha u^2  t}}.
\end{equation}
An expansion taking $m_i(k,s)$ small is thus valid if
\begin{equation}
 {u\over L} \gg   \sum_{\alpha:S_\alpha = S_i}  {{\overline\rho}_\alpha q_\alpha^2\over\epsilon  T}  {1\over 1 + {L^2\over 2D_\alpha u^2  t}}.
\end{equation}
The presence of the exponential in the integrand, and the fact that the integral vanishes for small $u$ means that the integral is dominated by $u\sim O(1)$ and thus we must have
\begin{equation}
 L \sum_{\alpha:S_\alpha = S_i}  {\overline{\rho}_\alpha q_\alpha^2\over\epsilon  T}  {1\over 1 + {L^2\over 2D_\alpha   t}} \ll 1.\label{ineq1}
\end{equation}
Now in the late time regime defined by $D_\alpha t/L^2 \gg 1$ the above reduces to the condition to be in the short range static regime
$Lm_i \ll 1$. 
In this regime the Laplace transform of the force is given by 
\begin{equation}
\tilde f_a \approx -{TA\over 8\pi s}  \int dk \ m_1(k,s) m_2(k,s)\exp(-2kL).
\end{equation}
In this form the Laplace transform for the time derivative of the force can be inverted to give
\begin{equation}
{d\over dt}\langle f(t)\rangle \approx -{TA\over 4\pi} \int dk \sum'_{\alpha\beta}{m_\alpha m_\beta
D_\alpha D_\beta k^2\over 
D_\beta -D_\alpha}(\exp(-2D_\alpha k^2t)- \exp(-2D_\beta k^2t))  \exp(-2kL).\label{int1} 
\end{equation}
the prime in the sum indicating that $\alpha$ and $\beta$ are species in different plates and we have defined $m_\alpha = {\overline\rho}_\alpha q_\alpha^2/\epsilon T$.  Fortunately the right hand side of Eq (\ref{int1}) can also be written as a temporal derivative and so we obtain
\begin{equation}
\langle f(t)\rangle \approx \langle f\rangle_{eq} +{TA\over 8\pi} \int dk \sum'_{\alpha\beta}{ m_\alpha m_\beta D_\alpha D_\beta \over 
D_\beta -D_\alpha}({1\over D_\alpha}\exp(-2D_\alpha k^2t)- {1\over D_\beta}\exp(-2D_\beta k^2t))  \exp(-2kL).
\end{equation}
The integral over $k$ can be expressed in terms of the complementary  error function
\begin{equation}
{\rm erfc}(u) = {2\over \sqrt{\pi}}\int_u^\infty du\ \exp(-u^2)
\end{equation}
as
\begin{equation}
\langle f(t)\rangle \approx \langle f\rangle_{eq} +{TA\over 16\sqrt{2\pi t}} \sum'_{\alpha\beta}{ m_\alpha m_\beta D_\alpha D_\beta \over 
D_\beta -D_\alpha}\left({1\over D_\alpha^{3\over 2}}\exp({L^2\over 2D_\alpha t}){\rm erfc}({L\over \sqrt{2D_\alpha t}})- {1\over D_\beta^{3\over 2}}\exp({L^2\over 2D_\beta t}){\rm erfc}({L\over \sqrt{2D_\beta t}})\right).
\end{equation}
The late time correction term is positive so the magnitude of the average  force approaches 
its equilibrium value from below with diffusive scaling. The appearance of diffusive scaling is interesting in itself. Even though $D_\alpha t$ represents the distance diffused by an ion of type $\alpha$, the ions diffuse in the plates, not across the gap between the plates and the Coulomb interaction is instantaneous !  The ions thus have to diffuse a distance of the order of the distance between the plates in order for the force to reach its equilibrium value, however this distance, in this limit, is less than the screening length. 

In the limit where $L/\sqrt{D_\alpha t} \ll 1$ we can use Taylor expansion of ${\rm ercf}(u)$,
\begin{equation}
{\rm erfc}(u) \approx 1-{2\over \sqrt{\pi}} u + O(u^2)
\end{equation}
to find
\begin{equation}
\langle f(t)\rangle \approx \langle f\rangle_{eq} +{TA\over 16\sqrt{2\pi t}} \sum'_{\alpha\beta}{ m_\alpha m_\beta D_\alpha D_\beta \over 
D_\beta -D_\alpha}\left({1\over D_\alpha^{3\over 2}}- {1\over D_\beta^{3\over 2}}\right).
\end{equation}
so the late time correction is independent of $L$ and decays as $1/\sqrt{t}$, this is reminiscent of the late time correction for Gaussian binary liquids where the force decays to its equilibrium value with a power law in time, independent of the plate separations \cite{deacout,gamcout}. 

\section{Late time evolution in the long range regime for the perfectly symmetric case}

The asymptotic inversion of the Laplace transform is quite subtle in the long range limit where $Lm_i \gg1$. In order to consider a system with only one intrinsic length scale and one intrinsic time scale we consider a perfectly symmetric system where all ions and anions have the same value of $m_\alpha=m_0$ and the same diffusion constant  $D_\alpha = D$.
Hence we consider the case where all the ions are cations are identical up to a change in the sign of their charge. 
As a consequence we can write
\begin{equation}
m_i(k,s) = {m\over  1+{ s\over 2 D k^2}},
\end{equation}
where $m = N_s m_0$ with $N_s$ the number of cations and anions in the (identical) plates. In this case 
we find
\begin{equation}
\tilde f_a = -{TA\over 2\pi s}  \int k^4dk  {m^2 D^2 \exp(-2kL)\over [s + 2D k^2 + mDk (1+\exp(-kL))][s + 2D k^2 + mDk(1- \exp(-kL))]},
\end{equation}
and the Laplace transform develops poles which are easy to find analytically and so it can be inverted to give
\begin{eqnarray}
\langle f(t)\rangle &=& -{TA\over 4\pi }  \int k^2dk \ m\exp(-kL) \left[ {1-\exp\left[ -Dtk (2k + m-m\exp(-kL))\right]\over  (2k + m-m\exp(-kL))}- \right.\nonumber \\
&&\left.{1-\exp\left[ - Dtk(2k + m+m\exp(-kL))\right]\over  (2k + m+m\exp(-kL))}\right].
\end{eqnarray}
From this we can easily verify the results for short time regime in this special case.  
As the final static result is known,  it is easiest to study the force dynamics via the temporal derivative 
\begin{equation}
{d\langle f(t)\rangle\over dt} = -{TADm\over 4\pi }  \int k^3dk \exp(-kL) \left(\exp\left[ -Dtk (2k + m-m\exp(-kL))\right]- \exp\left[ -Dtk (2k + m+m\exp(-kL))\right]\right).
\end{equation}
If we measure both the distance between the plates and the diffusion constant in terms of the screening length $1/m$, i.e. we write $L=\tilde L/m$ and $D = \tilde D/m^2$, then we find
\begin{equation}
{d\langle f(t)\rangle\over dt} = -{TA m^3\over 4\pi\tilde D^3 t^4 }  \int u^3du 
 \left(\exp\left[ -\tilde  g_-(u,\tilde D t,\alpha)\right]- \exp\left[ -g_+(u,\tilde Dt,\alpha)\right]\right).
\end{equation}
where 
\begin{equation}
g_\pm(u,\tilde D t,\alpha) = u({2u\over \tilde Dt}+\alpha + 1\pm \exp(-u\alpha)),
\end{equation}
and $\alpha = \tilde L/ \tilde Dt$.
In the late time limit $\tilde Dt \gg 1$  the functions $g_\pm$ simplify to
\begin{equation}
g_\pm(u,\alpha) = u(\alpha + 1\pm \exp(-u\alpha)),
\end{equation}
and so the temporal derivative of the force has the scaling form
\begin{equation}
{d\langle f(t)\rangle\over dt} = -{TA m^3\over 4\pi\tilde D^3 t^4 } Q(\alpha).
\end{equation}
We see that in this limit the behavior of the force evolution is not diffusive and is determined by the 
{\em ballistically} scaled parameter $\alpha = L/mDt$ rather than the diffusively scaled one $L^2/Dt$. 

In the limit $\alpha\to 0$ one can show
\begin{equation}
Q(\alpha) \approx 1/2\alpha^2
\end{equation}
which leads to
\begin{equation}
{d\langle f(t)\rangle\over dt} \approx -{TA m^3\over 8\pi\tilde D t^2 \tilde L^2}.
\end{equation}
Subsequently at late times
\begin{equation}
\langle f(t)\rangle \approx  -{TA\over 8\pi L^3}\left(\zeta(3) -{L \over  m D t  }\right).
\end{equation}
In the opposite limit $\alpha \to \infty$ we find asymptotically
\begin{equation}
Q(\alpha) \approx 48/(1+2\alpha)^5\approx 3/2\alpha^5,
\end{equation}
the first term being a much better approximation for $\alpha\sim 10$ for numerical verification of our analytical asymptotic estimates.  In the regime of $\alpha$ large we thus find
\begin{equation}
\langle f(t)\rangle \approx  -{3TA m^2 D^2 t^2\over 16\pi L^5}.
\end{equation}
This agrees with the general short time result given by Eq. (\ref{st}).

In this large distance regime we see that while the static behavior has a well defined universal  limit, independent of the microscopic details of the charges in the plates, the temporal behavior is strongly non-universal and depends on the parameters of the surface charges. 

\section{Conclusion}
We have studied the thermal fluctuation induced interaction between two plates that are modeled as 2D conductors containing mobile anions and cations. We analyzed the static limit as well as the dynamical approach to this limit. The attractive interaction is found to be not surprisingly a consequence of the build-up of correlations between the charge distributions on the two plates. Our results are however significantly different from those found between dielectric slabs studied in detail before \cite{dea2013}. While the static thermal Casimir interaction in the universal limit can also be obtained by approximating the interaction between dielectric slabs with diverging dielectric constants, the dynamics of the two systems is radically different and can not be extracted from this simplification. The late time relaxation of the van der Waals force between dielectric slabs is typically given by a simple exponential relaxation of the effective Hamaker coefficient (unless one invokes a power law distribution of dipole relaxation times) while the dynamics of the thermal Casimir effect in the model here has a much richer phenomenology. 

In fact, for plates with mobile anions and cations the behavior of the force is a complicated function of plate separation $L$ and time $t$. Several regimes were identified. In the short distance regime,  where effective charge fluctuations appear monopolar, the relaxation to the equilibrium force exhibits diffusive scaling.  In the large separation regime the static result is universal, and the dynamical behavior exhibits a ballistic-like scaling. Results obtained from the model A (non-conserved order parameter) dynamics for Gaussian binary mixtures exhibit a diffusive scaling in the evolution of the Casimir force to its equilibrium value \cite{gamcout,deacout}. However, the dynamics in this case occurs in the medium between the two plates, the only influence of the plates is to impose boundary conditions.  The appearance of the scaling variable $L^2/t$ is thus natural as the fluctuations of the order parameter in the $z$ direction between the plates obey a randomly (thermally) forced diffusion equation and relax over a region of size $L$. In the model considered here, nothing actually diffuses in the medium between the plates, all diffusion being confined within the plates and so in the {\sl living conductor } model the diffusive scaling is not as easy to anticipate and is quite unexpected on physical grounds.

Finally we note that the dynamics of the living conductor model is also a possible test bed for other non-equilibrium  studies of the Casimir effect, for instance one could model plates held at different temperatures by taking a different  amplitude of the thermal noise in each plate. One could also explore the effect of non-thermal driving on the dynamics of the  ions in the plates. The influence of hydrodynamics, due to the presence of a solvent, on the dynamics of the charges would also be interesting to examine. 

\section{Acknowledgment}
R.P. acknowledges support by the U.S. Department of Energy, Office of Basic Energy Sciences, Division of Materials Sciences and Engineering under Award No. DE- SC0008176.

\end{document}